\documentclass[twocolumn,prd,floatfix,preprintnumbers,a4paper,nofootinbib,superscriptaddress]{revtex4-1}

\usepackage{epsfig,graphics,graphicx}
\usepackage{amsmath,amssymb,mathtools,amsfonts}
\usepackage{bm,soul}
\usepackage[usenames,dvipsnames]{color}
\usepackage{amssymb,url,psfrag,times}
\usepackage{relsize}
\usepackage[varg]{txfonts}
\usepackage[colorlinks, pdfborder={0 0 0}]{hyperref}
\usepackage{lineno,verbatim}
\usepackage[utf8]{inputenc}
\usepackage{ulem}
\usepackage{xcolor}
\usepackage{dsfont}
\usepackage[utf8]{inputenc}

\usepackage[english]{babel} 
\usepackage{blindtext}

\definecolor{LinkColor}{rgb}{0, 0, 0.75}
\definecolor{CiteColor}{rgb}{0, 0, 0.75}
\definecolor{UrlColor}{rgb}{0, 0, 0.75}
\hypersetup{linkcolor=LinkColor}
\hypersetup{citecolor=CiteColor}
\hypersetup{urlcolor=UrlColor}
\usepackage{inputenc}
\usepackage{pgfplots}
\usepackage{tikz}
\usetikzlibrary{shapes,arrows}
\usetikzlibrary{positioning}
\usetikzlibrary{backgrounds}

\newcommand{\braket}[2]{ {\langle {#1} \, | \, {#2} \rangle} }
\newcommand{\bra}[1]{ \langle {#1} |  }
\newcommand{\ket}[1]{ | {#1} \rangle }

\definecolor{lightblue}{rgb}{.82,.88,0.95}
\definecolor{lightred}{rgb}{0.95,.86,0.86}
\definecolor{yellow}{rgb}{0.95,0.95,0.86}
\definecolor{green}{rgb}{.90,1,0.95}
\definecolor{lightpurple}{rgb}{.95,0.85,0.95}

\tikzset{
    vertex/.style = {
        circle,
        fill            = black,
        outer sep = 2pt,
        inner sep = 1pt,
    }
}

\def\gr#1{General Relativity#1
  (GR#1)\gdef\gr{GR}}

\def\gw#1{gravitational wave#1}

\def\grad#1{gravitational radiation#1}

\def\fig#1{Fig.~\ref{#1}}

\def\Eqn#1{Equation~(\ref{#1})}
\def\eqn#1{Eq.~(\ref{#1})}
\def\ceqn#1{Eq.~\ref{#1}}

\newcommand{\Eqnsa}[2]{Equations~(\ref{#1}) and (\ref{#2})}
\newcommand{\eqns}[2]{Eqs.~(\ref{#1}-\ref{#2})}
\newcommand{\eqnsa}[2]{Eqs.~(\ref{#1}) and (\ref{#2})}

\def\lal#1{LIGO Analysis Library#1
  (LAL#1)\gdef\lal{LAL}}
\def\nrda#1{\nr{} Data Analysis#1
  (NRDA#1)\gdef\nrda{NRDA}}
\def\tt#1{\textit{transverse--traceless}#1
  (TT#1)\gdef\tt{TT}}
\def\et#1{Einstein Telescope#1
  (ET#1)\gdef\et{ET}}
\def\ego#1{European Gravitational Observatory#1
  (EGO#1)\gdef\ego{EGO}}
\def\elisa#1{Evolved Laser Interferometer Space Antenna#1
  (eLISA#1)\gdef\elisa{eLISA}}
\def\ligo#1{Laser Interferometer Gravitational Wave Observatory#1
  (LIGO#1)\gdef\ligo{LIGO}}

\def\aligo#1{Advanced LIGO#1
  (Adv. LIGO#1)\gdef\aligo{Adv. LIGO}}
\def\snr#1{signal to noise ratio#1
  (SNR#1)\gdef\snr{SNR}}
\def\psd#1{power spectral density#1
  (PSD#1)\gdef\psd{PSD}}
\def\rom#1{reduced order model#1
  (ROM#1)\gdef\rom{ROM}}
\def\gatech#1{Georgia Institute of Technology#1
  (GaTech#1)\gdef\gatech{GaTech}}
\def\ffi#1{Fixed-Frequency Integration#1
  (FFI#1)\gdef\ffi{FFI}}
\def\sxs#1{Simulating Extreme Spacetimes#1
  (SXS#1)\gdef\sxs{SXS}}
\def\bam#1{Bifunctional Adaptive Mesh#1
  (BAM#1)\gdef\bam{BAM}}
\def\adm#1{Arnowitt-Deser-Misner
	(ADM#1)\gdef\adm{ADM}}
\def\frmse#1{Fractional Root-Mean Square Error
	(FRMSE)\gdef\frmse{FRMSE}}

\def\bh#1{black hole#1
 (BH#1)\gdef\bh{BH}}
\def\bbh#1{binary black hole#1
 (BBH#1)\gdef\bbh{BBH}}

\def\qnm#1{Quasi-Normal Mode#1
(QNM#1)\gdef\qnm{QNM}}
\def\eob#1{Effective One Body#1
  (EOB#1)\gdef\eob{EOB}}

\def\gw#1{gravitational wave#1}

\def\pn#1{Post-Newtonian#1
 (PN#1)\gdef\pn{PN}}
\def\pnl#1{post-Newtonian-like#1
  (PN-like#1)\gdef\pnl{PN-like}}

\def\nr{Numerical Relativity
 (NR)\gdef\nr{NR}}
\def\pt{\bh{} perturbation theory}

\def\rd{ringdown}

\def\pca#1{principle component analysis#1
  (PCA#1)\gdef\pca{PCA}}
\def\svd#1{Singular Value Decomposition#1
  (SVD#1)\gdef\svd{SVD}}

\def\adj#1{{#1}^{\dagger}}

\def\mcl{\mathcal{L}}
\def\mct{{\mathcal{T}}}

\def\lm{{{\ell m}}}
\def\lpm{{{\ell' m}}}
\def\l{{{\ell}}}

\def\lp{{{\ell'}}}

\newcommand{\brak}[2]{ \braket{#1}{#2} }
\newcommand{\cw}{\tilde{\omega}}

\def\gmvp#1{greedy-multivariate-polynomial#1
  (\texttt{GMVP}#1)\gdef\gmvp{\texttt{GMVP}}}
\def\gmvr#1{greedy-multivariate-rational#1
  (\texttt{GMVR}#1)\gdef\gmvr{\texttt{GMVR}}}

\makeatletter
\renewcommand*\env@matrix[1][*\c@MaxMatrixCols c]{%
  \hskip -\arraycolsep
  \let\@ifnextchar\new@ifnextchar
  \array{#1}}
\makeatother

\newcommand{\MITLIGO}{MIT Kavli Institute for Astrophysics and Space Research and LIGO Laboratory, 77 Massachusetts Avenue, Cambridge, MA 02139, USA}
\newcommand{\MITPhysicsKavli}{Department of Physics and MIT Kavli Institute, 77 Massachusetts Avenue, Cambridge, MA 02139, USA}
\newcommand{\Amsterdam}{Institute for High-Energy Physics, University of Amsterdam, Science Park 904, 1098 XH Amsterdam, The Netherlands}
\newcommand{\KCL}{King's  College  London,  Strand,  London  WC2R  2LS,  United Kingdom}

\begin{document}

\title{ Bi-orthogonal harmonics for the decomposition of gravitational radiation II: applications for extreme and comparable mass-ratio black hole binaries }

\author{L.\ London}
\email[]{lionel.london@kcl.ac.uk}
\affiliation{\KCL}
\affiliation{\Amsterdam}
\affiliation{\MITLIGO}

\author{S.\ A.\ Hughes}
\affiliation{\MITPhysicsKavli}

\begin{abstract}
	The estimation of a physical system's normal modes is a fundamental problem in physics. The quasi-normal modes of perturbed Kerr black holes, with their related spheroidal harmonics, are key examples, and have diverse applications in gravitational wave theory and data analysis. Recently, it has been shown that \textit{adjoint}-spheroidal harmonics and the related spheroidal multipole moments may be used to estimate the radiative modes of arbitrary sources. In this paper, we investigate whether spheroidal multipole moments, relative to their spherical harmonic counterparts, better approximate the underlying modes of binary black hole spacetimes. We begin with a brief introduction to adjoint-spheroidal harmonics. We then detail a rudimentary kind of spheroidal harmonic decomposition, as well as its generalization which simultaneously estimates pro- and retrograde moments.  Example applications to numerical waveforms from comparable and extreme mass-ratio binary black hole coalescences are provided. We discuss the morphology of related spheroidal moments during inspiral, merger, and ringdown. We conclude by discussing potential applications in gravitational wave theory and signal modeling. 
\end{abstract}

\date{\today}

\maketitle

%
%
\paragraph*{Introduction }--
%
%
~Each new \bbh{} detection provides new opportunities to compare the predictions of \gr{} with astrophysical data~\cite{LIGOScientific:2018mvr,LIGOScientific:2021djp,Mateu-Lucena:2021siq}. 
For current and future \gw{} detectors such as LIGO, KAGRA, Virgo and LISA, such comparisons are essential to the objectives of \gw{} science, from signal detection, to the estimation of source parameters, tests of \gr{}, and astrophysical interpretation~\cite{KAGRA:2013rdx}.  
In practice, each scientific objective is underpinned by our ability to accurately and efficiently represent the multipole moments of \grad{}~\cite{TheLIGOScientific:2016wfe,LIGOScientific:2016ebw}. 
While there is no unique definition of a \bbh{} system's radiative multipole moments, some definitions result in moments that closely resemble the system's natural modes, potentially simplifying signal representation, modeling and analysis~\cite{London:P1,London:2018gaq,London:2014cma,Holzegel:2013kna,Garcia-Quiros:2020qpx}. 
\par A motivation of Ref.~\cite{London:P1} (hereafter ``Paper I'') was to review, and perhaps improve upon, a common method for defining \gw{} multipole moments: decomposition with spin weighted $-2$ spherical harmonics. 
While the usefulness of the spin weighted spherical harmonics is derived from their completeness (i.e. their ability to exactly represent arbitrary \gw{} signals), and their orthogonality (meaning their multipole moments may be computed by projection, rather than e.g. fitting), their primary deficit is well known~\cite{London:2014cma,Kelly:2012nd,Garcia-Quiros:2020qpx,Ramos-Buades:2020noq}.
Spherical harmonics are closely related to the natural modes of \gw{} sources that have zero angular momentum.  
However, it is unclear if nature provides a mechanism to construct zero angular momentum objects and, to date, they have not been observed~\cite{Postnov:2019tmw,LIGOScientific:2018mvr,LIGOScientific:2021djp}. 
\par For this reason, Paper I considered the simplest generalization of spherical harmonics to spacetimes with angular momentum, namely, the spin weighted \textit{spheroidal} harmonics.
There, it was shown that, despite solving a non-classical\footnote{By ``non-classical'', we mean that the spheroidal harmonics relevant for e.g. Kerr black holes correspond to not one, but a countably infinite set of differential systems~\cite{London:P1}. This trait is shared by non-classical polynomials~\cite{Brezinski1991-cj}.} and non-hermitian eigen-problem, spheroidal harmonics are capable of exactly representing \textit{arbitrary} \gw{} signals.
It was also shown that spheroidal harmonics possess a non-trivial form of bi-orthogonality: rather than being orthogonal with themselves, a new sequence of special functions is required.
In Paper I, these functions are referred to as \textit{adjoint}-spheroidal harmonics because they are closely related to the eigenfunctions of the spheroidal differential equation's formal adjoint. 
\par For each spheroidal harmonic, there exists exactly one adjoint-spheroidal harmonic.
In the absence of angular momentum, the adjoint-spheroidal harmonics reduce to spherical harmonics, meaning that they can be efficiently and non-perturbatively computed using invertible linear operators.
%
%
A central result of Paper I is that spheroidal harmonics' use in defining \gw{} multipole moments causes the nullification of dominant nonphysical mode-mixing effects, resulting in a \textit{necessarily} simpler representation of each moment's amplitude and phase. 
\par In this \textit{Letter}, we survey potential applications of spheroidal harmonic decomposition to \gw{} theory. 
We begin with a brief review of the adjoint-spheroidal harmonics, followed by a description of how a standard change-of-basis approach allows existing sets of spherical harmonic moments to be directly converted into spheroidal ones. 
Here, we treat the angular momentum of the system in a very crude way: for \bbh{s} we define the spheroidal harmonics using the angular momentum of the remnant \bh{}~\cite{Mehta:2019wxm,Garcia-Quiros:2020qpx}.
This simple framing allows us to also consider a powerful generalization of spheroidal decomposition that simultaneously estimates pro- and retrograde mode content~\cite{Lim:2019xrb}. 
We then consider two numerical applications: first, the spheroidal harmonic decomposition of \gw{s} from a non-precessing extreme mass-ratio \bbh{} merger, and second, estimation of pro- and retrograde moments from a precessing comparable mass-ratio \bbh{} merger. 
We conclude with a brief discussion of possible future directions.
%
\paragraph*{Preliminaries }--
%
Gravitational wave theory is centrally interested in describing physical observables. Chief among them are the \gw{} polarizations, $h_+$ and $h_\times$~\cite{Blanchet:2013haa}.
It is common to consider both polarizations simultaneously by defining a complex valued strain, $h$, as
\begin{align}
	h \; = \; h_+ \; - \; i\,h_\times \; .
\end{align}
\begin{figure*}[htb] 
	\begin{tabular}{cc}
		\includegraphics[width=0.44\textwidth]{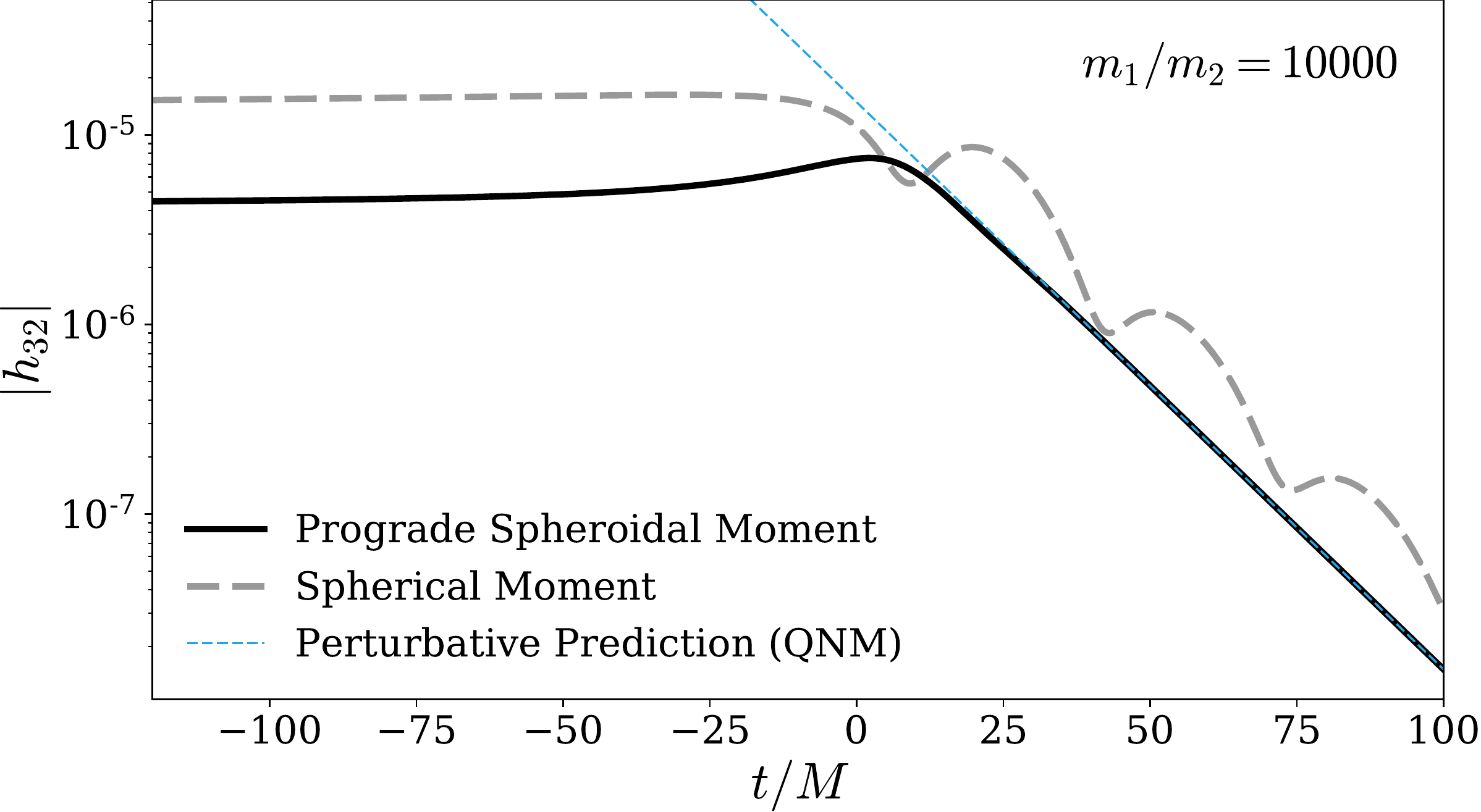} 
		&
		\includegraphics[width=0.44\textwidth]{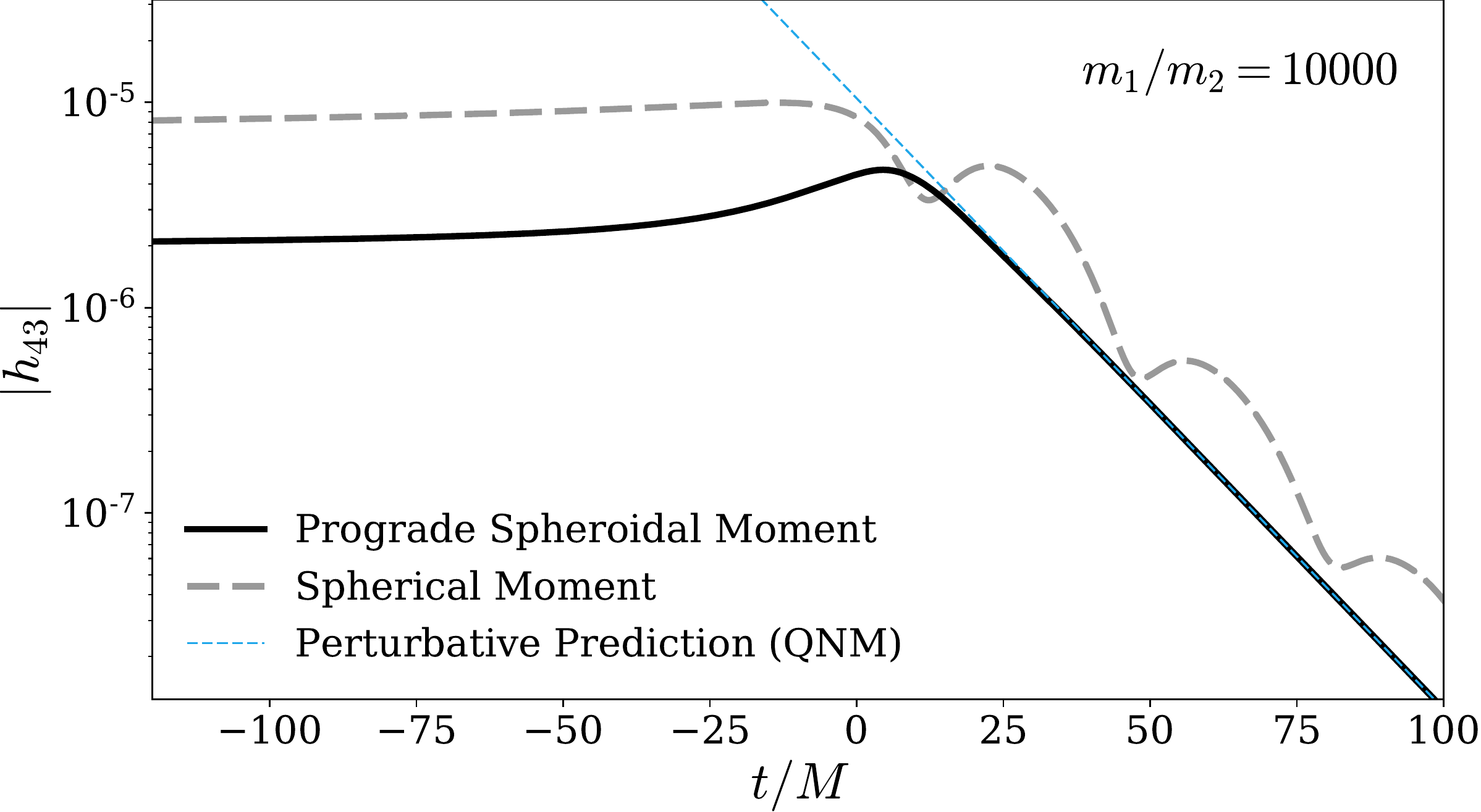} 
		\\
		\includegraphics[width=0.44\textwidth]{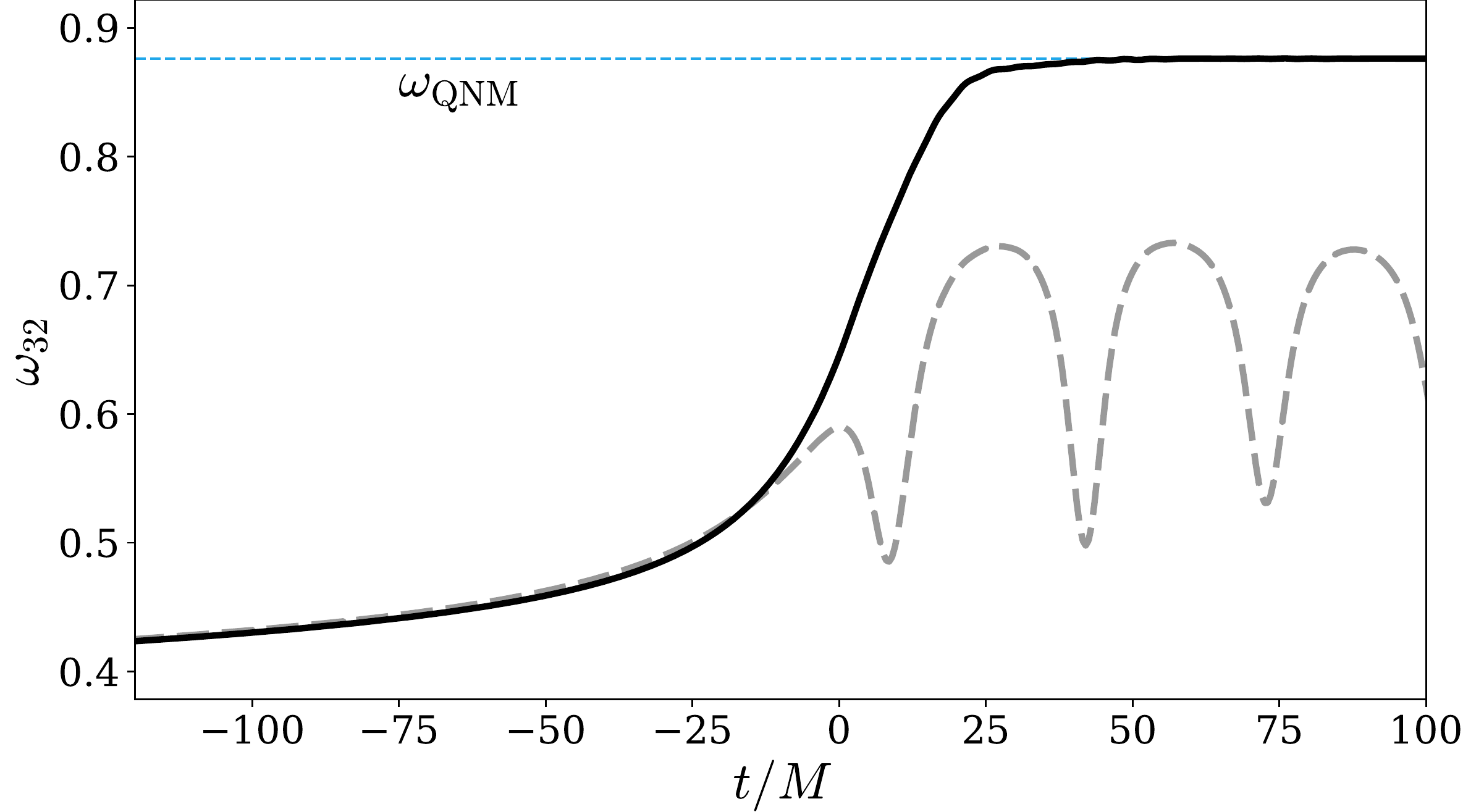} 
		&
		\includegraphics[width=0.44\textwidth]{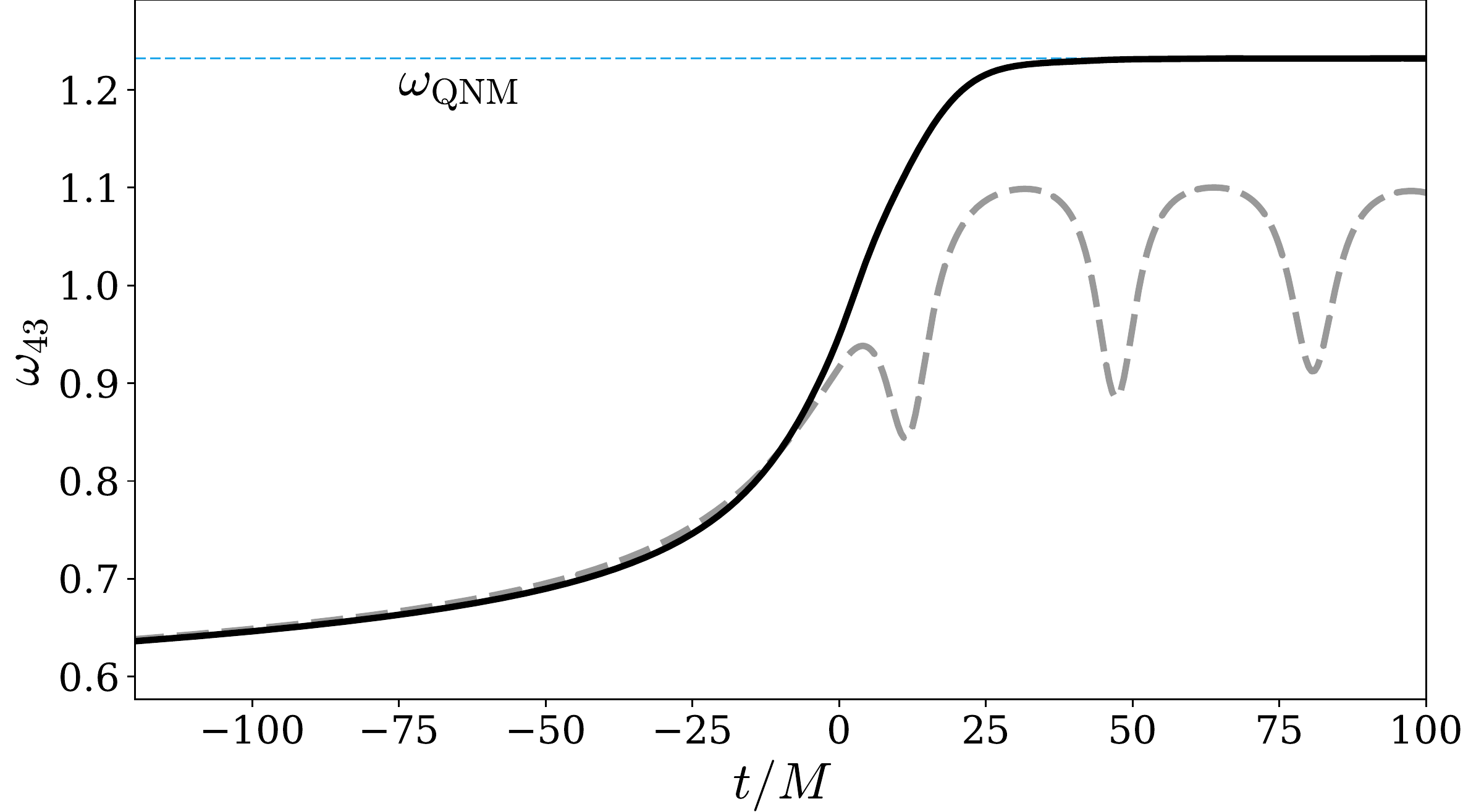} 
	\end{tabular}
	\caption{
		Example of spheroidal harmonic decomposition: radiative multipole moments from numerical binary black hole with $m_1/m_2=10000$, and a spin of $a_1=0.9$ on the larger body. Spheroidal moments are defined by $\brak{\tilde{S}_\lm}{h}$~(see \ceqn{hs0b}), and spherical ones by $\brak{\tilde{Y}_\lm}{h}$. Top panels compare multipole amplitudes. The perturbative predictions (thin dashed blue) are fits to the numerical data, where the long-linear slope is determined by single \bh{} perturbation theory (i.e. the dominant quasi-normal mode). The bottom panels compare multipole phase derivatives, $\omega_{\lm}=\partial_t \arg (h_\lm)$. Spheroidal harmonic multipole moments (black) are shown along side spherical harmonic moments (dashed grey), for the $(\ell,m)=(3,2)$ (left) and $(\ell,m)=(4,3)$ (right) moments. Here, $\omega_\mathrm{QNM}$ is determined only by \pt{}. Late time agreement between perturbative predictions and spheroidal moments signals that the correct physical modes are being estimated.
	}
	\label{fig:mixing_comparison}
\end{figure*}
In Paper I, it was shown that \gw{} polarizations from an arbitrary signal within \gr{} may be exactly equated with its spheroidal harmonic decomposition,
\begin{align}
	\label{hs}
	h(r,t,\theta,\phi) \; &= \; \frac{1}{r} \; \sum_{\ell,m} \, h^S_{\ell m}(t) \, {_{-2}}S_{\ell m}(\theta; \gamma_{\ell m } ) \, e^{i m \phi} \; , 
\end{align}
where $h^S_{\ell m}$ are the spheroidal multipole moments, $r$ is the \gw{}'s luminosity distance, $t$ is the observer's time coordinate, $\theta$ and $\phi$ are the usual spherical polar and azimuthal angles, defined in a source centered frame where the total angular momentum is along the $z$-direction, and $_{-2}S_{\ell m}$ are spin-weighted spheroidal harmonics.
In \eqn{hs}, $\ell$ and $m$ are the usual spherical harmonic indices, and $\gamma_\lm$ is an oblateness parameter~\cite{Berti:2005gp}.
If the source has mass $M$, and angular momentum $\vec{J}$, then the oblateness parameter is $\gamma_\lm = a \cw_{\ell m}$, where $a=|\vec{J}|/M$, and $\cw_\lm$ is an $M$ and $a$ dependent \qnm{} frequency of the spacetime~\cite{leaver85,Berti:2016lat,London:2018nxs}.
In what follows, we will focus on Kerr \bh{s}, for which $\gamma_\lm$ may defined in terms of the fundamental prograde \qnm{} frequencies\footnote{Note that in addition to the technical motivations for this choice discussed in Paper I, the proposed orthogonality of the radial Teukolsky functions in overtone label, $n$, (via a $3$D bilinear form) provides a natural setting for considering angular decomposition with subsets of \qnm{} with like $n$~\cite{Green,Sberna:2021eui}. In each framing, considering fixed $n$ subsets is supported by the non-degeneracy of spheroidal harmonic eigenvalues~\cite{London:P1}.}, $\gamma_\lm = a \cw_{\lm 0}$.
\par The choice of oblateness values, $\gamma_\lm$, fixes the decomposition space according to ${_{-2}}S_{\ell m}(\theta; \gamma_{\ell m } )$. 
In turn, this determines the adjoint-spheroidal functions: if $\mct$ is the linear invertible operator that transforms spherical harmonics into spheroidal harmonics, and we represent $_{-2}Y_\lm$ and $_{-2}S_\lm$ by their respective kets ($\ket{Y_\lm}$ and $\ket{S_\lm}$), then it can be shown that 
\begin{align}
	\label{T0}
	\mct \; = \; \sum_{\ell=2}^{\infty} \,\sum_{\lp =2}^{\infty} \, \ket{Y_\lm} \brak{Y_\lm}{S_\lpm} \bra{Y_\lpm} \;.
\end{align}
In \eqn{T0}, inner products are integrals over the usual solid angle. 
\Eqn{T0} may be used to non-perturbatively define the adjoint spheroidal harmonics in terms to their spherical harmonic expansion, 
\begin{align}
	\label{S0}
	\ket{\tilde{S}_{\lm}} \; = \; \adj{{\mct}^{-1}} \, \ket{Y_{\ell m}} \; .
\end{align}
In \eqn{S0}, $\dagger$ denotes adjugation. Given \eqns{T0}{S0}, $_{-2}\tilde{S}_{\lm}$ may be computed to machine precision via a finite-dimensional truncation of $\mct$~\cite{London:P1}.
Given $_{-2}\tilde{S}_{\lm}$, the spheroidal harmonic multipole moment of \eqn{hs} is defined via projection,
\begin{align}
	\label{hs0b}
	h^S_{\ell m} = \brak{\tilde{S}_\lm}{h}  =  \int_{0}^{2\pi} \int_{0}^{\pi} \, &{_{-2}}\tilde{S}^*_\lm(\theta;\gamma_\lm) \, e^{-im\phi} 
	\\ \nonumber
	&\times \, h(r,t,\theta,\phi) \, \sin(\theta)\, \mathrm{d}\theta \, \mathrm{d}\phi \, .
\end{align}
\par To gain intuition about the adjoint-spheroidal harmonics, it is illustrative to consider their perturbative expansion to linear order in $a$.
Directly applying the usual spheroidal harmonic perturbative expansion 
\begin{align}
	\label{s0}
	{_{-2}}{S}_{\ell m } &\approx Y_\lm \, + a \cw_{\lm 0}  c^{\l-1}_{\l m}Y_{\l-1,m}\, + a \cw_{\lm 0} c^{\l+1}_{\l m}Y_{\l+1,m} 
\end{align}
to \eqns{T0}{S0} yields, 
\begin{align}
	\label{s1}
	{_{-2}}\tilde{S}_{\ell m } \approx Y_\lm  - a \cw^*_{\l-1,m,0} c^{\l}_{\l-1,m}Y_{\l-1,m} - a \cw^*_{\l+1,m,0}c^{\l}_{\l+1,m}Y_{\l+1,m} \; .
\end{align}
In \eqnsa{s0}{s1}, $c^{\ell \pm 1}_{\ell m}$ are positive constants defined in Eqns. C13 and C14 of Ref.~\cite{OSullivan:2014ywd}, and $*$ denotes complex conjugation.
\par Comparing \eqn{s1} to \eqn{s0} reveals that unlike the spheroidal harmonics, the adjoint-spheroidal harmonics depend on not one, but multiple \qnm{} frequencies.
This reflects the fact that the each spheroidal harmonic with oblateness $\gamma_\lm$ is the eigenfunction of a differential operator, $\mcl(\gamma_\lm)$, where
\begin{align}
	\label{LSc}
	\mcl(\gamma_\lm)  =   u^2\gamma_\lm^2 -2su\gamma_\lm-\frac{(m+su)^2}{1-u^2}  + \partial_{u}(1-u^2)\partial_{u} \; .
\end{align}
In \eqn{LSc}, $s=-2$ and $u=\cos({\theta})$.
\par As there are countably infinite \qnm{} oblatenesses, $\gamma_\lm$, there are countably infinite operators $\mcl(\gamma_\lm)$ that must be taken into account when considering the orthogonality properties of the spheroidal harmonics. 
This concept underpins \eqn{T0}, and a key consequence of \eqns{T0}{S0} is that $_{-2}\tilde{S}_{\lm}$ and $_{-2}{S}_{\lm}$ are bi-orthogonal, meaning 
\begin{align}
	\label{B0}
	\brak{\tilde{S}_\lm}{S_\lpm}=\delta_{\ell \lp} / 2\pi \; .
\end{align}
%
\paragraph*{Spheroidal decomposition via change of basis}~--
%
It is common for the output of numerical simulations of \bbh{} mergers to store \gw{} data using spherical harmonic multipole moments, $\brak{Y_\lm}{h}$.
Thus, it is useful to note that, if given a set of {spherical} harmonic multipole moments, then \eqn{hs0b} need not be evaluated in order to compute the spheroidal ones. 
Instead, one may use \eqns{hs}{S0} to show that the matrix, $\hat{T}$, whose elements are, $\brak{Y_\lm}{S_\lpm}$, transforms vectors of spheroidal harmonic, $\vec{h_S}$, multipole moments into spherical ones, $\vec{h_Y}$,
\begin{align}
	\label{hy0}
	\vec{h_Y} \; = \; \hat{T} \, \vec{h_S} \; .
\end{align}
In \eqn{hy0}, $\vec{h_Y}$ has elements $\brak{Y_\lm}{h}$ where $m$ is fixed (due to the orthogonality of the complex exponentials in \eqn{hs}) and so only $\ell$ varies.
Similarly, $\vec{h_S}$ has elements $\brak{\tilde{S}_\lm}{h}$.
Thus the spheroidal harmonic multipole moments may be estimated via matrix inverse, 
\begin{align}
	\label{hy1}
	\vec{h_S} \; = \; \hat{T}^{-1} \, \vec{h_Y} \; .
\end{align}
Note that for the oblateness values considered here, $\hat{T}$ may be computed semi-analytically~\cite{OSullivan:2014ywd,leaver85,Cook:2014cta,London:P1}.
\par \Eqnsa{hs0b}{hy1} may be used to estimate the spheroidal harmonic moments of general \gw{} signals. 
As described in Paper I,  when applied to non-precessing \bbh{} systems, $h^S_\lm$ will have negligible mixing of different spheroidal modes during post-merger.  
%
%
In that setting, $h^S_\lm$ would, in principle, contain information about multiple overtone modes.
For precessing systems, $h^S_\lm$ will also contain information about different pro- and retrograde multipole moments, $h^{+S}_{\lm}$ and $h^{-S}_{\lm}$ respectively.
During ringdown, these moments correspond to perturbations that are pro- and retrograde with respect to the \bh{} horizon frequency.
\par A simple generalization of \eqn{hy1} allows $h^{+S}_{\lm}$ and $h^{-S}_{\lm}$ to be estimated simultaneously~\cite{Lim:2019xrb}.
However, this expansion of utility requires two additional assumptions that, unlike \eqn{hy1}, result in a form of least squares fitting that is not underpinned by the adjoint-spheroidal harmonics.
%
%
The first assumption is that each spheroidal moment is well modeled by 
\begin{align}
	\label{G0}
	 h^{+S}_{\lm} = A_\lm \exp(i \cw_{\lm 0} t) \;\text{ and }\; h^{-S}_{\lm} = A'_\lm  \exp(i \cw'_{\lm 0} t)  .
\end{align}
In \eqn{G0}, $A_\lm$ and $A'_\lm$ are complex valued, time dependent amplitudes, and $\cw_{\lm 0}$ and $\cw'_{\lm 0}$ are complex valued pro- and retrograde \qnm{} frequencies.
The second assumption is that $A_{\lm}(t)$ always evolves slowly relative to $i\cw_{\lm 0} t$, such that it may be treated as constant. 
Each assumption is explicitly compatible with late \qnm{} ringdown where one mode dominates.
Outside of ringdown, one may always find some $A_{\lm}(t)$ such that the first assumption holds.
The second assumption may, in general, break down due to e.g. very strong precession; however, departing from Ref.~\cite{Lim:2019xrb}, we note that the assumption is valid when the implicitly rotating source picture of \bbh{} merger holds~\cite{Baker:2008mj}.
Therefore the following generalization of \eqn{hy1} is particularly relevant for \bh{} \rd{}, but also expected to provide non-pathological results for most \bbh{} configurations.
{At the time of publication, what follows is to our knowledge the only known method for estimating pro- and retrograde multipole moments from inspiral and merger.}
\par The key idea is that to estimate both $h^{+S}_{\lm}$ and $h^{-S}_{\lm}$, one requires twice as much information as needed in spheroidal decomposition~(\ceqn{hy1}).
%
%
One robust way of obtaining this information is to consider $h^Y_\lm$, along with its first time derivative, $\partial_t {h}^Y_\lm$.
With \eqnsa{hs}{hy1} in mind, we must also explicitly consider the derivatives of each spheroidal moment. 
If we choose to treat the combined lists of $h^{+S}_{\lm}$ and $h^{-S}_{\lm}$ as we did $h^{S}_{\lm}$ in \eqnsa{hs}{hy1}, then the resulting algebraic system has the following schematic form,
\begin{figure}[t] 
	\begin{tabular}{c}
		\includegraphics[width=0.46\textwidth]{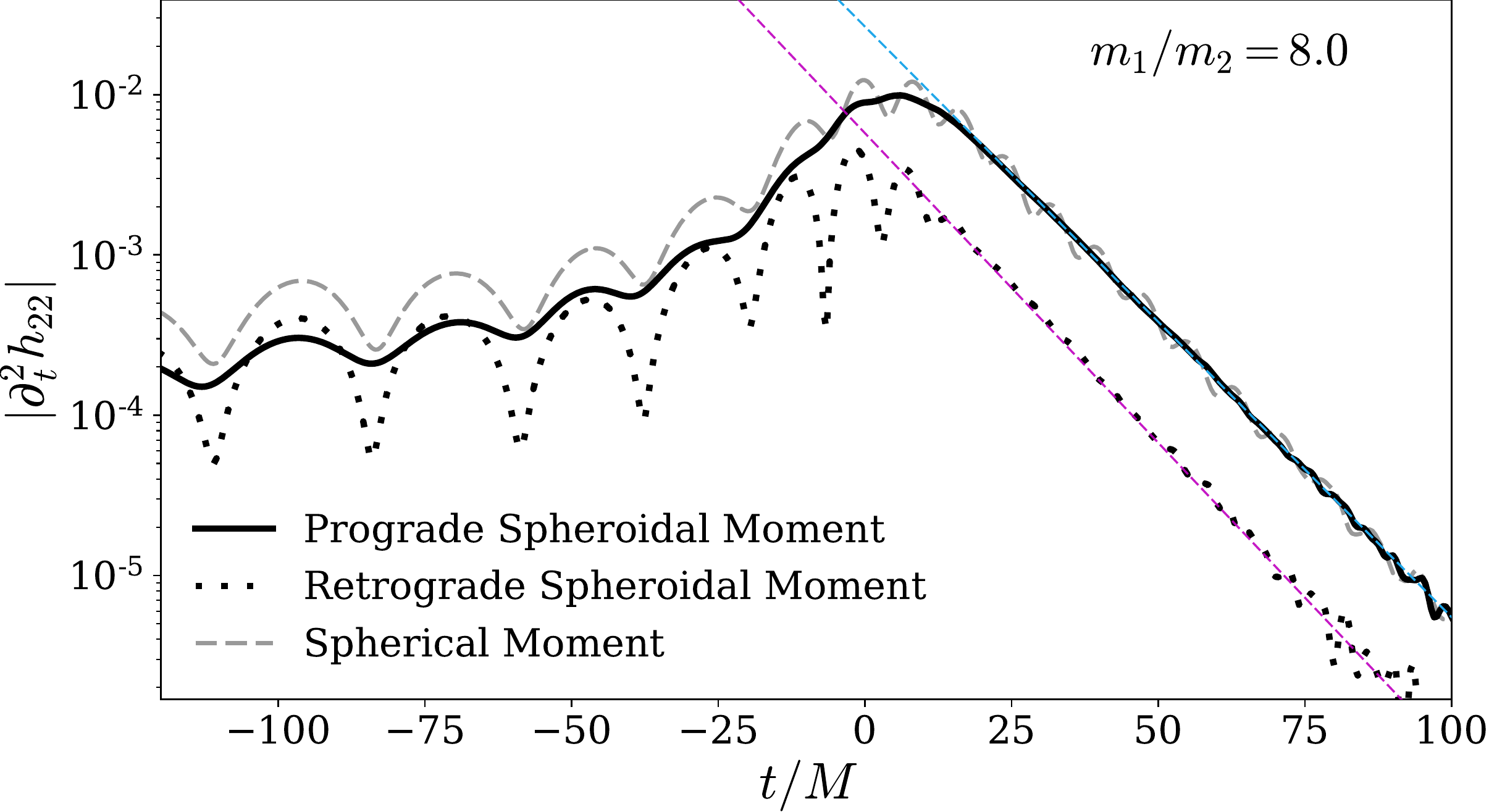} 
		\\ 
		\includegraphics[width=0.46\textwidth]{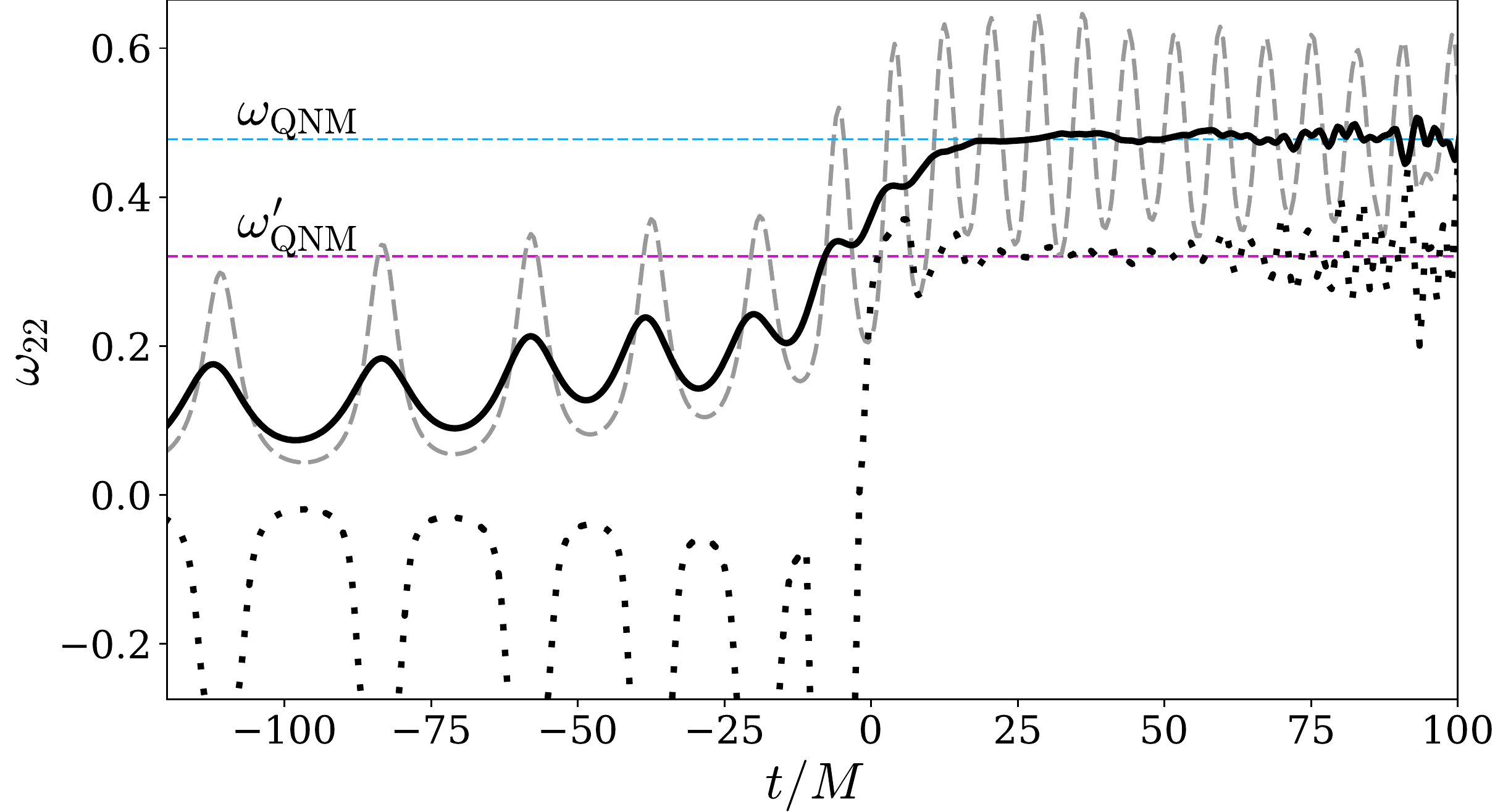} 
	\end{tabular}
	\caption{An example of spheroidal radiative moments from a mass ratio $8:1$ \bbh{} merger. The formatting of top and bottom panels is that same as that used in \fig{fig:mixing_comparison}. Retrograde frequencies (black dotted and magenta dashed lines) are negated for ease of presentation. The predicted \qnm{} retrograde frequency is $\omega'_\mathrm{QNM}$.
	}
	\label{fig:mixing_comparison_2a}
\end{figure}
\begin{align}
	\label{G1}
	\underbrace{
	\begin{pmatrix}[c]
		 h^Y_\lm 
		 \\ \hline 
		 \partial_t h^Y_\lm  \\
	 \end{pmatrix}
	}_{\mathlarger{\vec{H}_Y}}
	  = 
	 \underbrace{
	 \begin{pmatrix}[c|c]
		\brak{Y_\lm}{S_\lpm} & \brak{Y_\lm}{S'_\lpm} \\
		  \hline 
		  i\cw_{\lm 0} \brak{Y_\lm}{S_\lpm} & i\cw'_{\lm 0} \brak{Y_\lm}{S'_\lpm} \\
	  \end{pmatrix}
	  }_{\mathlarger{\hat{U}}}
	  \,
	 \underbrace{
	  \begin{pmatrix}[c]
		   h^{+S}_{\lm}
		   \\ \hline 
		   h^{-S}_{\lm}  \\
	   \end{pmatrix}
	   }_{\mathlarger{\vec{H}_S}}
\end{align}
In \eqn{G1}, each cell should be understood as the element of a vector or matrix whose rows and/or columns span values of $\ell$ (and/or $\lp$).
Each grouping of cells should be understood as vector and/or matrix concatenation, and each under-brace denotes the symbol we will use to represent each vector or matrix quantity. 
In the right-hand-side of \eqn{G1}, $S'_\lm$ are the spin weighted spheroidal harmonics defined with retrograde \qnm{} frequencies. 
\par \Eqn{G1} is the direct generalization of \eqn{hy0}.
We note that \eqn{G1} does not correspond to decomposition because, e.g., unlike \eqn{hy0}, in the limit of zero spacetime angular momentum, $\hat{U}$ does not reduce to the identity.
While in \eqn{hy0}, we wish to solve for $\vec{h}_S$, in \eqn{G1} we wish to solve for $\vec{H}_S$.
%
%
Just as in \eqn{hy1}, $\vec{H}_S$ (i.e. $h^{+S}_{\lm}$ and $h^{-S}_{\lm}$) may be determined by matrix inversion,
\begin{align}
	\label{G2}
	\vec{H}_S = \hat{U}^{-1} \vec{H}_Y \; .
\end{align}
%
%
In this way, \eqns{G0}{G2} allow the simultaneous estimation of pro- and retrograde moments through a kind of matrix least-squares fitting.
Note that neither \eqn{hy1} nor \eqn{G2} makes any assumption about the time domain behavior of spheroidal moments.
%
\paragraph*{Example application to extreme mass-ratio non-precessing binary black hole system}~--
%
In \fig{fig:mixing_comparison}, we show example spheroidal moments that result from applying \eqn{hy1} to a fiducial extreme mass-ratio binary black hole coalescence ($G=c=M=1$)~\cite{Apte:2019txp}.
The primary \bh{} of this system has a dimensionless spin of $a/M=0.9$, and the system's effective mass-ratio is $10,000:1$.
The secondary \bh{} is non-spinning, and follows an equatorial quasi-circular orbit.  
The purely prograde inspiral of this system means that only the prograde moment is excited. 
%
%
%
\paragraph*{Example application to comparable mass-ratio precessing binary black hole system}~--
%
In \fig{fig:mixing_comparison_2a}, we show example pro- and retrograde spheroidal moments that result from applying \eqn{G2} to the dominant quadrupole emission of a precessing \bbh{} coalescence~\cite{Hamilton:2021pkf}.
The primary \bh{} of this system has a dimensionless spin of $a/M=0.8$, and the system's mass-ratio is $8:1$.
The secondary \bh{} is non-spinning, and follows an initially quasi-circular orbit.  
The angle between the initial binary's orbital and spin angular momentum is $120^\circ$.
The non-zero value of this angle means that, e.g., the system's instantaneous orbital angular momentum precesses about $\vec{J}$, which itself varies slowly in direction and magnitude.
For consistency with the development of \eqn{G2}, this system's spherical multipole moments were transformed into a frame where $\vec{J}(t)$ always points along the $z$-axis~\cite{Hamilton:2021pkf,Ruiz:2007yx,Boyle:2011gg}.
The low-level data product for this simulation is the Weyl scalar, $\psi_4=\partial_t^2 \, h$~\cite{NP62,Ruiz:2007yx}.
In \fig{fig:mixing_comparison_2a}, we focus on this quantity to minimize the effect of data processing choices, and to more clearly display the pro- and retrograde spheroidal moments.
While results for this case are representative of many recent simulations of \bbh{} mergers, it was found that some older simulations (and multipole moments with $\ell>2$) are of insufficient accuracy to yield spheroidal moments of the quality shown here. 
%
\paragraph*{Discussion}~--
%
An ongoing challenge for \gw{} astronomy is to model \gw{} signals (numerically and analytically) in a way that is both precise and, ideally, explicitly encodes the physical principles particular to the astrophysical source and chosen theory of gravity.
Here, we have introduced and provided examples for two tools that may facilitate these goals: (\textit{i}) spheroidal harmonic decomposition, and (\textit{ii}) the matrix least-squares estimation of pro- and retrograde spheroidal multipole moments.
\par Like spin-weighted spherical harmonics, the spin-weighted spheroidal harmonics may be used in both numerical and analytic relativity to represent \gw{s} from arbitrary sources. 
Unlike spherical harmonics, the spheroidal harmonics are closely related to the modes of stationary spacetimes with angular momentum.
The techniques used to derive the adjoint-spheroidal harmonics may be applied to any scenario (within one theory, or across alternative theories of gravity) in which the field equations are separable~\cite{Berti:2009kk,London:P1,Noda:2022zgk}.
\par As seen in \fig{fig:mixing_comparison}, spheroidal harmonic decomposition may find use in representing \gw{s} from non-precessing \bbh{} systems: For the non-quadrupolar moments, the decomposition results in (\textit{a}) post-merger ($t/M>0$) waveforms that are naturally consistent with \pt{}, (\textit{b}) merger waveforms (near $t/M=0$) that are single-peaked, reminiscent of the quadrupole moments, and (\textit{c}) late inspiral ($t/M<0$) waveforms that appear to be qualitatively different from their spherical counterparts.
Analytic and numerical analyses of each difference, (\textit{a})-(\textit{c}), may be of future interest. 
\par Like spherical harmonics, the spin weighted spheroidal harmonics cannot explicitly determine pro- and retrograde multipole moments.
%
%
For this, versions of the matrix least-squares method encapsulated by \eqn{G2} may find use in the analysis of simulated \gw{s} from precessing \bbh{} systems.
In \fig{fig:mixing_comparison_2a} we see for the first time pro- and retrograde contributions to the multipole asymmetry in comparable mass-ratio \gw{} data, implying that this generalization of spheroidal decomposition may find use in, e.g., \gw{} signal modeling, where precession's imprints on radiative multipole moments are still being refined for use in current and future detection scenarios~\cite{Hamilton:2021pkf}. 
\par Note that like spheroidal decomposition, this fitting method makes no strong assumption about the time domain behavior of \gw{} signals; instead, it is assumed that the different multipole moments primarily depend on the inner-products within $\hat{U}$~(\ceqn{G1}). 
Thus agreement with expected time domain behavior evidences consistency between the different spheroidal moments that are encoded within each spherical one.
This consistency is related to the known inner-product ratio test for \bh{} \rd{}~\cite{London:2014cma,Kelly:2012nd}.
A matter of ongoing investigation is whether \eqn{G2} may be further extended to self-consistently estimate the presence of \bh{} overtones (see e.g. Refs.~\cite{leaver85,Berti:2016lat,London:2014cma,Bhagwat:2019dtm}).
%
%
\par Finally, we note that the results presented here imply at least two salient questions for future work:
(\textit{Q1}) Here we have used a very rudimentary form of spheroidal decomposition were the spacetime angular momentum is held fixed -- is it possible to generalize this to account for non-stationary angular momentum?
And lastly, (\textit{Q2}) given that the adjoint-spheroidal harmonics may be motivated using the angular Teukolsky equation~(See \ceqn{LSc} and e.g. Ref.~\cite{Press:1973zz}), do similar arguments apply to Teukolsky's radial equation, given that they are both generalized spheroidal (Heun-type) equations~\cite{leaver86,Cook:2014cta}?
\paragraph*{Acknowledgments --}
%
Work on this problem was supported at Massachusetts Institute of Technology~(MIT) by National Science Foundation Grant No. PHY-1707549 as well as support from MIT’s School of Science and Department of Physics. Lionel London was also supported at the University of Amsterdam by the GRAPPA Prize Fellowship, and at King's College London by the Royal Society University Research Fellowship, Award No. URF{\textbackslash}R1{\textbackslash}11451. Many thanks are extended to Mark Hannam for use of the numerical waveform presented in \fig{fig:mixing_comparison_2a}.

\definecolor{UrlColor}{rgb}{0.75, 0, 0}
\bibliographystyle{apsrev4-1}
\bibliography{p2.bib}
\end{document}